\documentclass[oupdraft]{bio}
\usepackage{url}
\usepackage{enumerate}
\usepackage{graphicx,verbatim,array,multicol, palatino}
\usepackage{amsthm, amsmath, amssymb,  setspace}
\usepackage{xcolor,colortbl}
\usepackage{color}
\usepackage{mathtools}
\usepackage{indentfirst}

\definecolor{red}{rgb}{1,0,0}
\definecolor{blue}{rgb}{0,0,1}
\definecolor{green}{rgb}{0,0.6,0.4}

\def\b{\boldsymbol}

\newcommand{\Mod}[1]{\ (\text{mod}\ #1)}
\DeclarePairedDelimiter\ceil{\lceil}{\rceil}
\DeclarePairedDelimiter\floor{\lfloor}{\rfloor}

\history{Received XXXXX;
revised XXXXX;
accepted for publication XXXXX}

\begin{document}

\title{Inferring Mobility Measures from GPS Traces with Missing Data}

\author{IAN BARNETT$^\ast$ \\[4pt]
\textit{Department of Biostatistics, University of Pennsylvania, 423 Guardian Drive, Philadelphia, PA 19104} \\[2pt]
JUKKA-PEKKA ONNELA \\[4pt]
\textit{Department of Biostatistics, Harvard University, 677 Huntington Avenue, Boston, MA 02115}
{ibarnett@pennmedicine.upenn.edu}}

\markboth%
{I. Barnett and J.P. Onnela}
{Inferring Mobility Measures from GPS Traces with Missing Data}

\maketitle

\footnotetext{To whom correspondence should be addressed.}

\begin{abstract}
{With increasing availability of smartphones with GPS capabilities, large-scale studies relating individual-level mobility patterns to a wide variety of patient-centered outcomes, from mood disorders to surgical recovery, are becoming a reality. Similar past studies have been small in scale and have provided wearable GPS devices to subjects. These devices typically collect mobility traces continuously without significant gaps in the data, and consequently the problem of data missingness has been safely ignored. Leveraging subjects' own smartphones makes it possible to scale up and extend the duration of these types of studies, but at the same time introduces a substantial challenge: to preserve a smartphone's battery, GPS can be active only for a small portion of the time, frequently less than $10\%$, leading to a tremendous missing data problem. We introduce a principled statistical approach, based on weighted resampling of the observed data, to impute the missing mobility traces, which we then summarize using different mobility measures. We compare the strengths of our approach to linear interpolation, a popular approach for dealing with missing data, both analytically and through simulation of missingness for empirical data. We conclude that our imputation approach better mirrors human mobility both theoretically and over a sample of GPS mobility traces from 182 individuals in the Geolife data set, where, relative to linear interpolation, imputation resulted in a 10-fold reduction in the error averaged across all mobility features.}
{Imputation, mHealth, Missing data, GPS, Mobility, Precision medicine}
\end{abstract}

\section{Introduction}
\label{sec1}
The Global Positioning System (GPS) is a navigation system that uses a device's distance from a number of satellites in orbit to determine the location of the device. GPS has a wide range of applications. For example, in transportation GPS has been used to complement self-report surveys on travel activity \citep{chapman2007integrating,stopher2007assessing,zhou2003analysis,shen2014review}. One study instrumented participants with both a GPS receiver and an accelerometer and showed that combining both data sources improved the prediction of a person's mode of activity \citep{troped2008prediction}. Another study outfitted children with GPS devices and showed that their travel behavior was different when they were accompanied by an adult as opposed to when they were on their own \citep{mackett2007setting}.

Another burgeoning area of application is mobile health and the application of GPS to social and behavioral research in a wide variety of contexts \citep{wolf2010gps}. For example, GPS devices on older ($>63$ years of age) care-recipients were used to show that caregiver burden was negatively correlated with the amount of time a care-recipient spent walking per day \citep{werner2012caregiving}. Two studies found that mobility measures extracted from GPS data were correlated with depressive symptom severity \citep{saeb2015mobile,canzian2015trajectories}, and another study was able to predict changes of state for bipolar patients with $80.8\%$ accuracy \citep{gruenerbl2014using}. Amongst pregnant women at risk for perinatal depression, women with a larger radius of travel were found to have milder depressive symptoms than the women with more severe symptoms \citep{faherty2017movement}. Following spine surgery, GPS tracking was used to monitor patient recovery and found that increased mobility corresponded with a successful recovery \citep{yair2011assessing}. GPS has also shown promise at keeping track of wandering dementia patients \citep{miskelly2005electronic} as well as at monitoring the mobility of patients with Alzheimer's disease \citep{shoval2008use}. By combining pollution measurements from air samples and a person's GPS trace, exposure levels at the individual-level were calculated \citep{phillips2001use}. Several behavioral traits measured passively through the smartphone use of schizophrenia patients were found to be associated with self-reported measures of mental health \citep{wang2016crosscheck}.

In most of the above studies, GPS data was collected from study participants either through a study-provided smartphone or wearable GPS receiver. While data collected in this way will have minimal missingness, there are three distinct disadvantages to supplying the GPS device. Firstly, this study design is not scalable because it is expensive to provide GPS devices to a large number of participants. Secondly, adherence to wearable devices typically declines sharply after a few months \citep{AARPref}, making long-term longitudinal studies less feasible. Thirdly, the data may be biased in unpredictable ways due to the interference that can arise by introducing of a new device into a participant's life \citep{ainsworth2013comparison}. All of these shortcomings can be avoided by taking advantage of built-in GPS devices in the smartphones of participants. In 2017, $80\%$ of U.S. adults owned smartphones, up from $35\%$ in 2011 \citep{smith2015us,cassagnol2017smartphone}, and this number is expected to continue to increase. Anonymized call detail records (CDRs), resulting from mobile phone communication events, have been used to study both social networks \citep{onnela2007structure} and mobility patterns \citep{gonzalez2008understanding} at scale, and analysis and modeling of CDRs to different purposes has since then become an active field of its own \citep{blondel2015survey}. However, for the purposes of inferring mobility metrics, these data are quite limited as the location of the person is only available at the level of the cell tower used to transmit the event, and even this is only available at the time of communication (either via calls or text messages). Smartphone-based mobility traces from GPS are therefore more precise both spatially and temporally and, importantly, making it possible to link the smartphone data with individual-level covariates, which in the context of \textit{digital phenotyping}, which we have previously defined as the ``moment-by-moment quantification of the individual-level human phenotype \textit{in situ} using data from personal digital devices'' \citep{onnela2016harnessing,torous2016new}, could range from simple demographic variables to fMRI imaging or genome sequencing data.

To make use of the ubiquity of smartphones in biomedical research settings, we have developed an open source research platform for digital phenotyping called Beiwe that includes customizable iOS and Android smartphone apps that, among other features, can record a phone's GPS trace using user specified sampling scheme \citep{torous2016new}. For a smartphone app to be scalable and enable long-term data collection, it must not impose too much of a drain on the phone's battery. Study adherence, which in this context means not uninstalling the app during the study, could be in jeopardy if the participant notices a significant drop in the battery life. Of all the current smartphone sensors, GPS is the most expensive with regards to battery usage \citep{miller2012smartphone}. To lower the strain on the battery, the Beiwe platform records GPS over short intervals, called on-cycles, between gaps of periods of inactivity, called off-cycles. The platform enables users to specify the length of on-cycles and off-cycles at will; for example, the on-period might be 2 minutes and the off-period 10 minutes. However, all of these off-periods lead to a large portion of the GPS trace being missing.

To the best of our knowledge, there is currently no principled method of handling continuous missing GPS data at the individual level \citep{krenn2011use,jankowska2015framework}. There have been methods developed that avoid using the coordinate pairs of GPS data by representing significant locations as nodes in a network \citep{liao2007learning}, and there are methods that require the trajectories of large samples of individuals to construct road networks \citep{li2016knowledge}. Studies that model trajectories at the individual-level so far have either ignored missing data \citep{canzian2015trajectories}, or have used linear interpolation assuming travel at constant $1$ m/s over the missing interval \citep{shin2007human,rhee2011levy}. With the rise of research in digital phenotyping large scale research studies aiming to measure patient-centered outcomes and behavioral phenotypes in naturalistic settings over long periods of time are becoming more prevalent, necessitating the development of statistical methods that properly account for missingness. Here we introduce a statistical approach for imputing the missing trajectories present in a mobility trace that attempts to simulate human mobility patterns at the individual level. The properties of this approach are compared analytically to linear interpolation, as well as across high-frequency complete-data mobility traces from 182 individuals in the Geolife data set \citep{zheng2009mining,zheng2008understanding,zheng2010geolife}. Compared to linear interpolation, our imputation approach offers significant reductions in error relative to the ground truth in the estimation of a wide variety of mobility measures, with a 10-fold improvement in the average error.

\section{Methods}
\label{sec2}

\subsection{Mapping longitude and latitude to a 2D plane}
\label{subsec21}

While raw GPS data consists of a sequence of longitude and latitude coordinates that trace a person's location on the surface of the Earth, most mobility metrics are computed for data in a 2D Euclidean plane, requiring a transformation of coordinates as the first step. Because of the differences in geometry, there will always be distortion when mapping the surface of a 3D sphere to a 2D plane, although the distortion is smaller the smaller the region of the sphere's surface being mapped. People typically do not travel far enough on a daily basis for this projection to greatly distort their mobility traces. For the purposes of extracting mobility measures, a person's mobility trace can be mapped to a 2D plane on an individual basis, as opposed to using a univerisal projection such as the Mercator projection \citep{maling2013coordinate}. By allowing each person their own projection, distortion can be minimized by selecting the projection best suited for each individual. We detail these individualized projections here.

\begin{figure}
\centerline{\includegraphics[scale=.5,clip=true,trim=0 2.5cm 0 0 ]{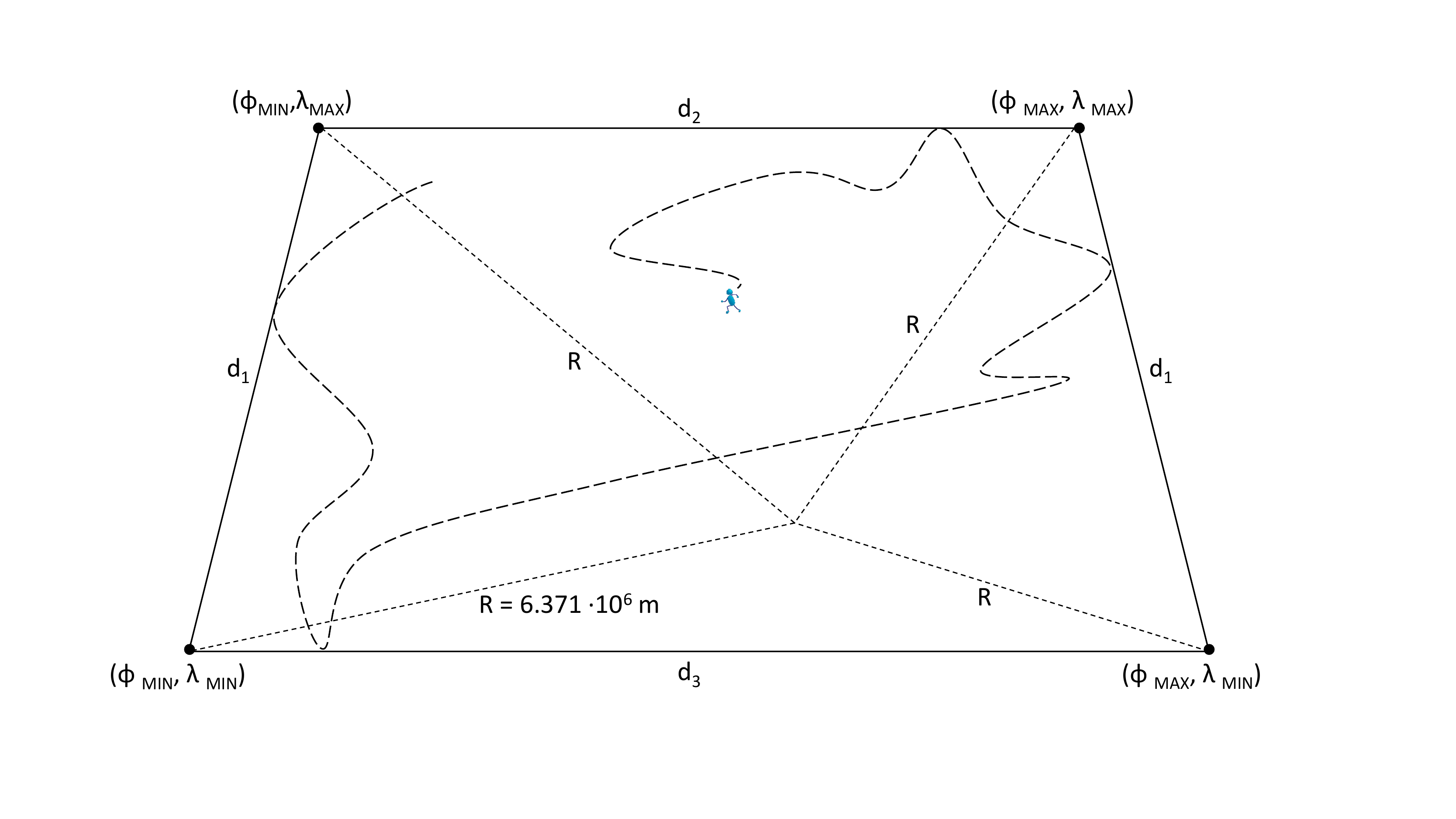}}
\caption{\textbf{Schematic of longitude-latitude projection to X-Y plane.} The isosceles trapezoid contains the projection of the mobility trace for a particular individual. In the northern hemisphere $d_2 < d_3$ while in the southern hemisphere $d_2>d_3$. The long dashed curve represents a person's example mobility trace.} \label{ProjectionSchematic}
\end{figure}

Consider a person's mobility trace where we let $\lambda_{\min}$, $\lambda_{\max}$, $\phi_{\min}$, and $\phi_{\max}$ be the minimum and maximum latitude and longitude attained over the study period, respectively. By projecting the region bounded by the points ($\phi_{\min}$,$\lambda_{\min}$), ($\phi_{\max}$,$\lambda_{\min}$), ($\phi_{\min}$,$\lambda_{\max}$), and ($\phi_{\max}$,$\lambda_{\max}$) onto an isosceles trapezoid, the distortion of the projection is greatly reduced (see Figure \ref{ProjectionSchematic}). To map a specific $(\phi,\lambda)$ coordinate to the X-Y plane, let $w_{\lambda} = \frac{\lambda-\lambda_{\min}}{\lambda_{\max}-\lambda_{\min}}$, $w_{\phi} = \frac{\phi-\phi_{\min}}{\phi_{\max}-\phi_{\min}}$, $d_1 = (\lambda_{\max}-\lambda_{\min})\cdot R$, $d_2 = (\phi_{\max}-\phi_{\min})\cdot R\cdot \sin(\pi/2-\lambda_{\max})$, and $d_3 = (\phi_{\max}-\phi_{\min})\cdot R\cdot \sin(\pi/2-\lambda_{\min})$, where $R = 6.371 \cdot 10^{6}$ m represents the Earth's radius in meters. Then the corresponding $(x,y)$ pair is:
$$\begin{array}{rl}
x = & w_1 \left(\frac{d_3-d_2}{2}\right) + w_2\left\{d_3(1-w_1)+d_2w_1\right\},\\
y = & w_1 d_1 \sin\left\{\cos^{-1}\left(\frac{d_3-d_1}{2d_1}\right)\right\}.\\
\end{array}$$
As a reference point, we assign the origin to ($\phi_{\min}$,$\lambda_{\min}$).

\subsection{Notation and model}

First, a person's GPS latitude and longitude coordinates are transformed to 2D plane coordinates using the projection detailed in \ref{subsec21}. According to the rectangular method \citep{shin2007human}, the data are next converted into a mobility trace defined by a sequence of flights, corresponding to segments of linear movements, and pauses, corresponding to periods of time where a person does not move. Also, if a missing interval is flanked by two pauses at the same location (situated within 50 meters of one another), the missing interval is assumed to be a longer pause at the same location.

Suppose a person's mobility trace begins at time $t_0$ at projected coordinates $(x_0,y_0)$. The mobility trace is modeled as a sequence of $n$ events, where an event is either a flight of straight-line movement, or a pause. Let $\Delta^x_i$ be the horizontal displacement of the $i$th event, $\Delta^y_i$ be the vertical displacement of the $i$th event, and $\Delta^t_i$ be the duration of the $i$th event. The time of the $i$th event is $t_i$ while the location at the start of the $i$th event is $(x_i,y_i)$. Letting $\b{z}_i=(x_i,y_i,t_i)$,
$${\b z}_i = {\b z}_0 + \sum_{j=1}^{i-1}(\Delta^x_j,\Delta^y_j,\Delta^t_j).$$
If the $i$th event is unobserved or missing, then $m_i=1$. Due to the battery strain that GPS imposes, on smartphones GPS can only be activated for regularly scheduled short intervals (e.g. 2 minutes) of time, with large gaps (e.g. 10 minutes) between collection periods. This scheduled missingness occurs independent of a person's mobility and therefore can be classified as missing completely at random (MCAR)\citep{little2002statistical}, meaning that $\mbox{pr}(m_i=1|{\b Z}) =\mbox{pr}(m_i=1)$, where ${\b Z}=(X,Y,T)$ represents the random variables for location and time. While this scheduled missingness will undoubtedly account for the largest percentage of missing data in a person's GPS trace, some GPS data may be missing not at random (MNAR) as related to a person's mobility, such as powering off the phone, being inside a tall building, or geographical features inhibiting satellite connection, and this type of missing data is not accounted for here.

We model the event displacements as
$$(\Delta^x_i,\Delta^y_i,\Delta^t_i) \sim B_iF^{(f)}(\cdot|{\b Z}={\b z}_i) + (1-B_i)F^{(p)}(\cdot|{\b Z}={\b z}_i)$$
where $F^{(f)}$: $\mathbb{R}^3\rightarrow \mathbb{R}$ and $F^{(p)}$: $\mathbb{R}^3\rightarrow \mathbb{R}$ are the distribution functions for flights and pauses, respectively, $B_i=1$ when the $i$th event is a flight and $B_i=0$ when the $i$th event is a pause, with  $B_i \sim \mbox{Bernoulli}(p_i)$. The probability of the $i$th event being a flight instead of a pause is $p_i = B_{i-1}\Psi({\b z}_i)+(1-B_{i-1})$ where $\Psi$: $\mathbb{R}^3\rightarrow \mathbb{R}$ is the probability of observing a flight conditional on the previous event being a flight. The reason that $p_i$ is dependent on $B_{i-1}$ is because two consecutive pauses are impossible by definition, as they would simply combine to count as one longer pause at the same location. This forces $B_i=1$ conditional on $B_{i-1}=0$. Note that $F^{(f)}$ and $F^{(p)}$ are distribution functions conditional on ${\b Z}$ while $\Psi$ is not. This conditioning allows the full distribution functions of flights and pauses to change with ${\b Z}$.

\subsection{Continuity assumptions}
 
Because $F^{(f)}$, $F^{(p)}$, and $\Psi$ are unknown, we must rely on empirical  estimates of these functions from the observed data. With the goal of resampling from observed events, or hot-deck imputation, we must make several continuity-like assumptions on $F^{(f)}$, $F^{(p)}$, and $\Psi$ in order to enable local resampling from observed events to impute missing events. We assume $\Psi$ is continuous, and that $\forall {\b z} \in \mathbb{R}^3$ and for every $\epsilon>0$, there exists $\delta>0$ such that
\begin{align} 
\left\lVert F^{(f)}(\cdot|{\b Z}={\b z}+{\b \delta})-F^{(f)}(\cdot|{\b Z}={\b z})\right\lVert_{\infty} \leq \epsilon \label{fconteq1} \\
\left\lVert F^{(p)}(\cdot|{\b Z}={\b z}+{\b \delta})-F^{(p)}(\cdot|{\b Z}={\b z})\right\lVert_{\infty} \leq \epsilon \label{fconteq2}
\end{align}
for all $0\leq\delta_x<\delta$, $0\leq\delta_y<\delta$, and $0\leq\delta_t<\delta$, where ${\b \delta}=(\delta_x,\delta_y,\delta_t)$. This condition ensures that the distribution of flights and pauses are similar locally with respect to location and time.

Let $E_f = \{i\in\{1,...,n\}: B_i=1\}$ be the indices of flights with $n_f=|E_f|$. Suppose we wish to resample from the observed set of flights to impute a trajectory at some new time and location ${\b z}_{\mbox{new}}=(x_{\mbox{new}},y_{\mbox{new}},t_{\mbox{new}})$. The empirical distribution, giving $w_k(\cdot)$ weight to the $k$th flight, is 
\begin{align}
\hat{F}^{(f)}\left\{{\b \Delta}=(\Delta^{(x)},\Delta^{(y)},\Delta^{(t)})|{\b Z}={\b z}_{\mbox{new}}\right\} = \frac{\sum_{k \in E_F}w_k({\b z}_{\mbox{new}}) I_{\{{\Delta}_k^{(x)}<{\Delta^{(x)}},{\Delta}_k^{(y)}<{\Delta^{(y)}},{\Delta}_k^{(t)}<{\Delta^{(t)}}\}}}{\sum_{k \in E_F}w_k({\b z}_{\mbox{new}}) }. \nonumber
\end{align}

 In addition to the continuity assumption of equation \eqref{fconteq1}, $\epsilon = \epsilon(\delta)$ where $\epsilon(\cdot)$ is non-decreasing and $\epsilon(0)=0$, we consider the first two moments of the asymptotic distribution of the empirical distribution function:
\begin{align}
&\left|E\left[\sqrt{n_f}\left(F^{(f)}({\b \Delta}|{\b Z}={\b z}_{new})-\hat{F}^{(f)}({\b \Delta}|{\b Z}={\b z}_{new})\right)\right]\right|  \nonumber\\
 &=\left|\sqrt{n_f} \frac{\sum_{k \in E_F} w_k({\b z}_{new})\left(F^{(f)}({\b \Delta}|{\b Z}={\b z}_{new})- F^{(f)}({\b \Delta}|{\b Z}={\b z}_k)\right)}{\sum_{k \in E_F}w_k({\b z}_{new})}\right|  \nonumber\\
&\leq \sqrt{n_f} \frac{\sum_{k \in E_F} w_k({\b z}_{new})\epsilon\left(\lVert{\b z}_{new}-{\b z}_k\rVert_{\infty}\right)}{\sum_{k \in E_F}w_k({\b z}_{new})} \label{expbound}
\end{align}
\begin{align}
&\mbox{Var}\left[\sqrt{n_f}\left(F^{(f)}({\b \Delta}|{\b Z}={\b z}_{new})-\hat{F}^{(f)}({\b \Delta}|{\b Z}={\b z}_{new})\right)\right]  \nonumber\\
 &=n_f\frac{\sum_{k \in E_F} w^2_k({\b z}_{new})F^{(f)}({\b \Delta}|{\b Z}={\b z}_k)\left(1- F^{(f)}({\b \Delta}|{\b Z}={\b z}_k)\right)}{\left(\sum_{k \in E_F}w_k({\b z}_{new})\right)^2}.  \label{vareq}
\end{align}
If the distribution function of flights is independent of time and location, then $F^{(f)}(\cdot|{\b Z}) = F^{(f)}(\cdot)$, and so resampling can be performed as in the case of an independent and identically distributed sample by letting $w_i(\cdot) = 1/n_{f}$ for each event. In this case $\epsilon(\delta)=0 \;\; \forall \;\delta>0$, so the expectation in Equation \eqref{expbound} reduces to $0$ and the variance in Equation \eqref{vareq} simplifies to the binomial variance, $F^{(f)}({\b \Delta})(1-F^{(f)}({\b \Delta}))$. 

However, it is unlikely that the distribution of flights or pauses is identically distributed across all times and locations. In this case, the empirical distribution function will be biased, with a bound for the magnitude of this bias specified in Equation \eqref{expbound}. The bound for this bias is minimized by giving higher weight to events that are closer in time and location to the new event ${\b z}_{\mbox{new}}$. This can be achieved by specifying $ w_k({\b z}_{new})$ so that it is inversely related to $\lVert{\b z}_{\mbox{new}}-{\b z}_k\rVert_{\infty}$. In this extreme, letting $w_k({\b z}_{\mbox{new}})=I_{\{{\b z}_k={\b z}_{\mbox{new}}\}}$ would eliminate the bias completely, but is impractical is it as it wouild assign a weight of $0$ every other observed event. To minimize the variance in Equation \eqref{vareq}, the weights are spread out equally across all events $w_k({\b z}_{\mbox{new}})=1/n_{f}$, but this will lead to an inflated bias. Instead, a balance must be achieved when selecting weights so that both the bias and variance of the empirical distribution function are kept low. This can be done by selecting weights from a unimodal function centered on ${\b z}_{\mbox{new}}$. To this effort, we choose a $t$-distribution function with $\nu$ degrees of freedom in order to allow for both spread and kurtosis to be controlled as tuning parameters. The same principles in selecting weights can be applied to the empirical approximations of $F^{(p)}(\cdot)$ and $\Psi(\cdot)$. The empirical approximation to $\Psi(\cdot)$ is
\begin{align}
\hat{\Psi}({\b z}_{\mbox{new}}) = \frac{\sum_{j=2}^n B_{j-1}B_j w_j({\b z}_{\mbox{new}}) }{\sum_{j=2}^nB_{j-1}w_j({\b z}_{\mbox{new}}) }. \nonumber
\end{align}

In order to improve resampling further, in addition to the continuity assumptions of Equations \eqref{fconteq1} and \eqref{fconteq2} we also consider several potentially realistic assumptions on human mobility:
\begin{enumerate}[i.]
\item \textit{Temporally local} (TL) weights: Events close in time tend to have similar mobility patterns.
$$\begin{array}{rll}
F^{(f)}(\cdot|{\b Z}) = & F^{(f)}(\cdot|T) & \\
F^{(p)}(\cdot|{\b Z}) = & F^{(p)}(\cdot|T) & \\
\Psi({\b Z}) = & \Psi(x,y,T) & \forall x,y\in\mathbb{R}.\\
\end{array}$$
Resampling weights corresponding to this assumption are:
\begin{align}
w_j({\b z}_{new})=  \psi_{\nu}\left(c\cdot (t_{new}-t_{j})\right), \nonumber
\end{align}
where $\psi_{\nu}(\cdot)$ is the $t$-distribution density function with $\nu$ degrees of freedom and $c$ is a scaling constant.
\item \textit{Geographically local} (GL) weights: Events close in space tend to have similar mobility patterns. 
$$\begin{array}{rll}
F^{(f)}(\cdot|{\b Z}) = & F^{(f)}(\cdot|X,Y) &\\
F^{(p)}(\cdot|{\b Z}) = & F^{(p)}(\cdot|X,Y) & \\
\Psi({\b Z}) = & \Psi(X,Y,t) & \forall t \in\mathbb{R}.\\
\end{array}$$
Resampling weights corresponding to this assumption are:
\begin{align}
w_j({\b z}_{\mbox{new}})= \psi_{\nu}\left(c\cdot \sqrt{(x_{\mbox{new}}-x_j)^2+(y_{\mbox{new}}-y_j)^2}\right). \nonumber
\end{align}
\item \textit{Geographically local with circadian routine} (GLC) weights: Events close in space and close in the time of day have similar mobility patterns. Considering time to be measured in hours:
$$\begin{array}{rll}
F^{(f)}(\cdot|X,Y,T=t) = &F^{(f)}(\cdot|X,Y,T=t+24k ) & \forall k\in\mathbb{Z}, \forall t \in\mathbb{R} \\
F^{(p)}(\cdot|X,Y,T=t) = &F^{(p)}(\cdot|X,Y,T=t+24k) & \forall k\in\mathbb{Z}, \forall t \in\mathbb{R} \\
\Psi(X,Y,T) = &\Psi(X,Y,T+k\cdot24  \mbox{ hours}) & \forall k\in\mathbb{Z}. \\
\end{array}$$
Letting $s$ represent 24 hours (in the units of time of $t$) and letting $c_1$ and $c_2$ be the scaling constants, the resampling weights corresponding to this assumption are:
\begin{align}
w_j({\b z}_{new}) = &\psi_{\nu}\left\{c_1\cdot \sqrt{(x_{new}-x_j)^2+(y_{new}-y_j)^2}\right\} \nonumber\\
&\cdot \psi_{\nu}\left[c_2 \cdot \min\left\{|t_{new}-t_j|\Mod{s},s-|t_{new}-t_j|\Mod{s}\right\}\right]. \nonumber
\end{align}
\end{enumerate}
 
There are reasonable arguments for any of the TL, GL, or GLC assumptions. Human mobility patterns may be a function of location. For example, people may be more stationary when at home than when outside the home. In this case the GL assumption would be able to ensure low values of $\Psi({\b z}_i)$ when $(x_i,y_i)$ are the coordinates of home a person's home. The GLC assumption adds to this a circadian component which would be better at recovering information about, say, a regular commute. The TL assumption would do well to model bursty human movement where flights tend to occur in bunches over time. This assumption captures what is likely the most general and robust pattern of human behavior.

\subsection{Imputing missing trajectories}

Our approach to dealing with missing data is to impute missing flight and pause events by resampling from observed events over each missing interval or sequence of events. A person's true Cartesian location after projection at time $t$ is ${\b L}(t) = (L_x(t),L_y(t))$. Consider a period of missing data in a mobility trace that starts at time $\tau_s$ and ends at time $\tau_f$.  ${\b L}(t)$ is unobserved over the time interval $(\tau_0,\tau_1)$ with ${\b L}(\tau_0)$ and ${\b L}(\tau_1)$ both being known. We aim to closely approximate ${\b L}(t)$ over the missing interval $(\tau_0,\tau_1)$ by borrowing information from the observed mobility trace outside this interval. Previous approaches either ignore missing data altogether \citep{canzian2015trajectories} or have used linear interpolation (LI) between ${\b L}(\tau_0)$ and ${\b L}(\tau_1)$, which essentially amounts to connecting the dots at ${\b L}(\tau_0)$ and ${\b L}(\tau_1)$ with a straight line. More precisely, linear interpolation estimates ${\b L}(t)$ with $\tilde{{\b L}}(t) = \frac{\tau_1-t}{\tau_1-\tau_0}{\bf L}(\tau_0) +\frac{t-\tau_0}{\tau_1-\tau_0}{\b L}(\tau_1)$ on the interval $(\tau_0,\tau_1)$, and this same linear trajectory is assumed at some pre-specified constant velocity, such as 1 m/s, from ${\b L}(\tau_0)$ to ${\b L}(\tau_1)$ \citep{shin2007human,rhee2011levy}. While such a simple model of human mobility may be suitable for near-complete data with scarce missingness, for smartphone GPS data with substantial degree of missingness, a more careful treatment is required.

To simulate a trajectory at any given time point $t$, we first simulate if a flight occurs (as opposed to a pause) with probability $\Psi\left(L_x(t),L_y(t),t\right)$. If a flight is determined to occur, then a flight is sampled from the empirical distribution function $\hat{F}^{(f)}(\cdot|{\b Z}=(L_x(t),L_y(t),t))$. If a pause is determined to occur, then a pause is sampled from $\hat{F}^{(p)}(\cdot|{\b Z}=(L_x(t),L_y(t),t))$. Letting the displacement for the $b$th simulated consecutive event be denoted as ${\b \Delta}_{(b)}$, events are simulated until number of steps $q$ is reached such that $\sum_{b=1}^{q+1} \Delta_{(b)}^{t} \geq \tau_1-\tau_0$ is true, at which point the process is terminated and ${\b \Delta}_{(q)}$ is declared as the displacement of the final simulated event of the imputed trajectory over the missing interval.

Let $N(t) = \max\{k: \tau_0+\sum_{b=1}^k \Delta_{(b)}^t \leq t \}$ be the counting process for the number of simulated events elapsed by time $t$ and let $t_{(b)}$ be the time of the $b$th event in the simulated trajectory. The simulated trajectory, used as an estimator for ${\b L}(t)$, bridged so it starts at ${\b L}(\tau_0)$ and ends at ${\b L}(\tau_1)$ is:
\begin{align}
&\hat{{\b L}}(t) = \frac{t_{(N(t)+1)}-t}{t_{(N(t)+1)}-t_{(N(t))}}{\b G}(t_{(N(t))}) + \frac{t-t_{(N(t))}}{t_{(N(t)+1)}-t_{(N(t))}}{\b G}(t_{(N(t)+1)}) \nonumber
\end{align}
where
\begin{align}
&{\b G}(t) = \frac{\tau_1-t}{\tau_1-\tau_0}\left\{{\b L}(\tau_0)+\sum_{b=1}^{N(t)}\left(\Delta^x_{(b)},\Delta^y_{(b)}\right)\right\}+\frac{t-\tau_0}{\tau_1-\tau_0}{\b L}(\tau_1). \nonumber
\end{align}
This bridging ensures that the flights retain the property of being straight lines, whereas bridging $G(t)$ directly would lead to curvature.

Above we described our process for simulating a person's trajectory over an interval $(\tau_0,\tau_1)$ of missing data. A person's full mobility trace is likely to have multiple missing intervals, and this same approach can be applied equally to each missing interval. Imputing over the gaps in a person's mobility trace in this fashion introduces variability, so repeated imputations over the same missing intervals are likely to produce variable trajectories. Mobility metrics, such as radius of gyration and distance travelled, will also vary with each imputed mobility trace. As a result, repeated imputations can be used to provide confidence bounds for any mobility metric of interest to account for the uncertainty that results from data missingness and subsequent imputation. After repeating simulating the same trajectories $B$ times and calculating the desired mobility metrics each time, the $\alpha/2 \cdot B$ and the $(1-\alpha/2) \cdot B$ ordered values form the lower and upper confidence bounds of an $\alpha$-level confidence interval, respectively.

\section{Results}
\label{sec3}

\subsection{Analytical treatment of the expected gap between a mobility trace and its surrogates} \label{AnalyticalSection}

Here we consider a model for a person's mobility trace and then compare analytically the performance of our approach to linear interpolation in the ability to approximate the true trajectory. Consider a mobility trace with no pauses where each flight has the same arbitrary duration of one unit of time and where the $x$ displacement is independently distributed from the $y$ displacement of a flight. For the $x$ and $y$ flight displacements, let the expectations be functions of $t$, $\mu_x(t)$ and $\mu_y(t)$, and let the variances be constant, $\sigma_x^2$ and $\sigma_y^2$, respectively. We assume each flight to be independent. Though such stringent independence assumptions would lead unrealistic mobility traces, the analytic results that can be derived based on this model will provide some insight into how the extent of missingness is related to the accuracy of trajectories we simulate over the missing periods.

For $t \in (0,n)$ the mobility trace is ${\b L}(t) = (t-\floor{t})(\Delta_{\ceil{t}}^{x},\Delta_{\ceil{t}}^{y}) + \sum_{i \leq t}(\Delta_{i}^{x},\Delta_{i}^{y})$. Without loss of generality we assume ${\b L}(0)=(0,0)$. If instead of continuous time we consider discrete time $t \in \{0,1,\dots,n\}$, this simplifies to ${\b L}(t)=\sum_{i=1}^t(\Delta_{i}^{x},\Delta_{i}^{y})$. Though ${\b L}(t)$ represents the actual trajectory, assume that the time period $(0,n)$ represents a period of missingness, and thus a period where ${\b L}(t)$ is unobserved. We use a  simplified version of our simulated trajectory estimate $\hat{{\b L}}(t)$, which has been bridged so that $\hat{{\b L}}(0)={\b L}(0)$ and $\hat{{\b L}}(n)={\b L}(n)$, where we only consider integer-valued $t$: $\hat{{\b L}}(t) =\frac{t}{n}\left\{{\b L}(n)-\sum_{i =1}^n (\Delta_{(i)}^{x},\Delta_{(i)}^{y})\right\}+ \sum_{i =1}^t (\Delta_{(i)}^{x},\Delta_{(i)}^{y})$. Here $\Delta_{(i)}^{x}$ and $\Delta_{(i)}^{y}$ represent the $x$ and $y$ displacements of the $i$th resampled flight, and are assumed to be independent from and distributed the same as $\Delta_{i}^{x}$ and $\Delta_{i}^{y}$, respectively. Compare this to linear interpolation, which in this context simplifies to $\tilde{{\b L}}(t) = \frac{t}{n}{\b L}(n)$. Ideally, the simulated trajectory $\hat{{\b L}}(t)$ and the linearly interpolated trajectory $\tilde{{\b L}}(t)$ should be `close' to the true but unobserved trajectory ${\b L}(t)$. To measure closeness, we examine the average squared distance between the ${\b L}(t)$ and $\hat{{\b L}}(t)$ across $t \in \{0,1,\dots,n\}$, or $\frac{1}{n+1}\sum_{t=0}^nE\left[\left\lVert {\b L}(t)-\hat{{\b L}}(t) \right\rVert^2\right]$. We then do the same to compare the closeness of ${\b L}(t)$ and $\tilde{{\b L}}(t)$. We seek to answer the question of how the length of the period of missingness, in this case $n$, relates to the accuracy of the surrogate trajectories $\hat{{\b L}}(t)$ and $\tilde{{\b L}}(t)$ used to replace the unobserved true trajectory ${\b L}(t)$.

\begin{figure} 
\centerline{\includegraphics[scale=1]{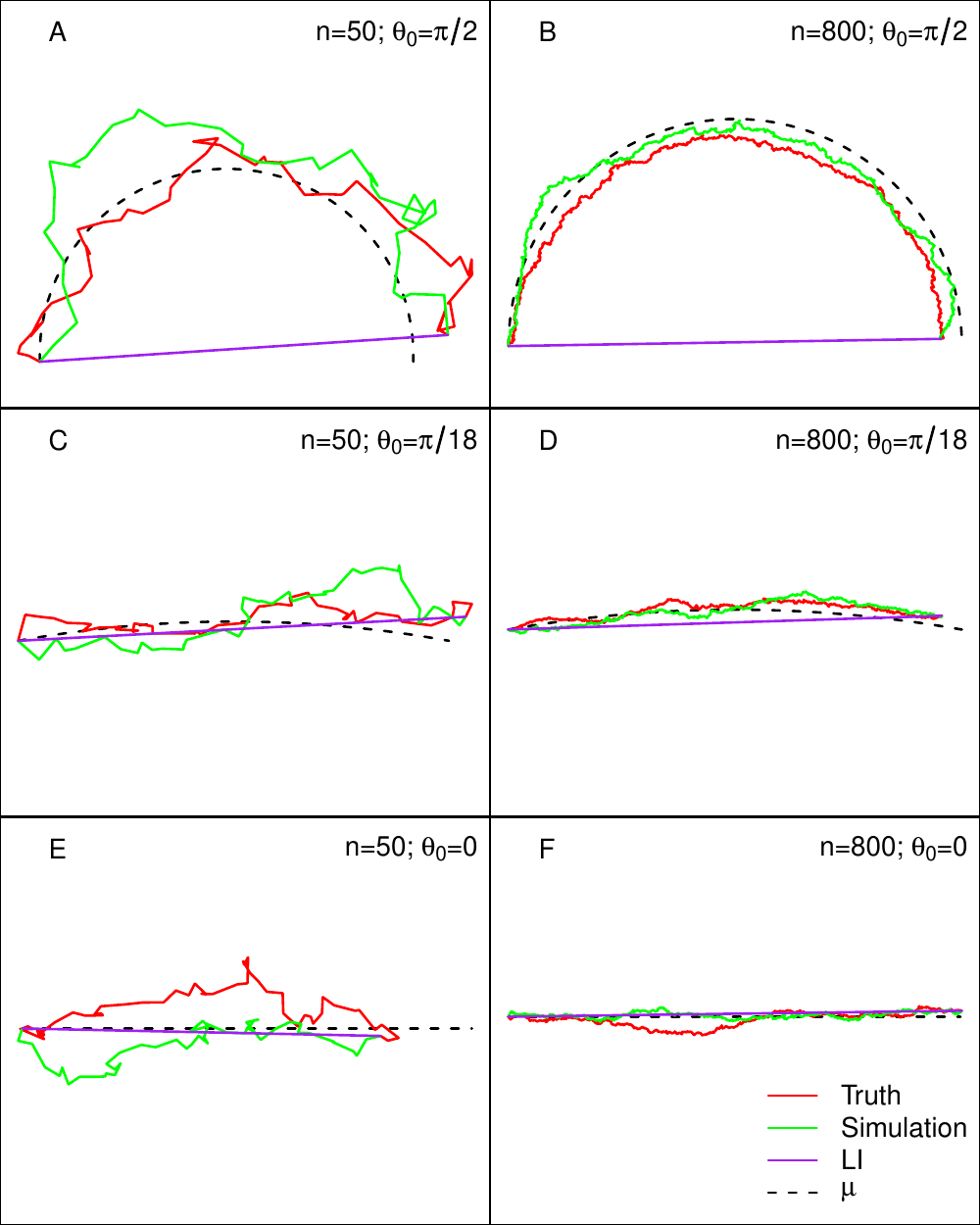}}
\caption{\textbf{Theoretical unobserved trajectories and their surrogates.} Trajectories are generated according to the theoretical model of Section \ref{AnalyticalSection}. Panels (A), (C), and (E) represent a shorter period of missingness ($n=50$) while panels (B), (D), and (F) represent a longer period of missingness ($n=800$). The red trajectory represents a person's true, unobserved mobility trace over an interval of $n$ units of time, and the dashed line represents its expected trajectory. It is assumed that the location immediately before and immediately after this interval are observed and known. The green trajectory represents one simulated instance using our approach, while the purple trajectory represents linear interpolation as a means for imputing the missing gap. The $\theta_0$ represents the starting angle above the $x$-axis of the mean trajectory. Linear interpolation is best when the expected trajectory is a straight line, but only the simulation approach is robust to curvature.} \label{SemiCircleTrajExFig}
\end{figure}

We consider a family of trajectories to allow for varying degrees of curvature in ${\b L}(t)$. For a fixed $\theta_0 \in [0,\pi/2]$ we consider the mean displacements of the flight at time $t \in \{0,1,\dots,n-1\}$ to be $\mu_x(t) = \sqrt{d}\cos\left(\theta_0-\frac{2\theta_0t}{n-1}\right)$ and $\mu_y(t) = \sqrt{d}\sin\left(\theta_0-\frac{2\theta_0t}{n-1}\right)$, where $d$ is the expected distance of a flight. Under this model, $\theta_0=0$ corresponds to a straight trajectory whereas $\theta_0=\pi/2$ corresponds to a semicircular trajectory (see Figure \ref{SemiCircleTrajExFig}). We investigate how close the simulated trajectory $\hat{{\b L}}(t)$ is to the actual trajectory ${\b L}(t)$: $E\left[\left\lVert {\b L}(t)-\hat{{\b L}}(t) \right\rVert^2\right] = 2t(1-\frac{t}{n})(\sigma_x^2+\sigma_y^2)$.
By averaging this quantity across all $t$ in the missing interval we arrive at:
\begin{equation}\label{EHt}
\frac{1}{n+1}\sum_{t=0}^nE\left[\left\lVert {\b L}(t)-\hat{{\b L}}(t) \right\rVert^2\right] = \left(\frac{n-1}{3}\right)(\sigma_x^2+\sigma_y^2).
\end{equation}

An analogous calculation can be performed to see how close the linearly interpolated trajectory  $\tilde{{\b L}}(t)$ is to ${\b L}(t)$: $E\left[\left\lVert {\b L}(t)-\tilde{{\b L}}(t) \right\rVert^2\right] = t\left(1-\frac{t}{n}\right)(\sigma_x^2+\sigma_y^2) + M(t)$,
where
$$
\begin{array}{rl}
M(t)=&\sum_{l\in\{x,y\}}\bigg[ \sum_i^t \mu_l(i)^2+\frac{t^2}{n^2}\sum_{i=1}^n\sum_{j=1}^n\mu_l(i)\mu_l(j)-\frac{2}{n}\sum_{i=1}^t\sum_{j=1}^n\mu_l(i)\mu_l(j) \\
&+ 2\sum_{i<j}^t\bigg\{\mu_l(i)\mu_l(j)-\frac{1}{n}\sum_{k=1}^n\mu_l(k)(\mu_l(i)+\mu_l(j))\bigg\}\bigg]
\end{array}$$
is a function of the $\mu(i)$. The derivation is left for the Supplementary Materials. Again we average across all time points in the missing interval to arrive at:
\begin{align}\label{EFt}
& \frac{1}{n+1}\sum_{t=0}^nE\left[\left\lVert {\b L}(t)-\tilde{{\b L}}(t) \right\rVert^2\right] = \left(\frac{n-1}{6}\right)(\sigma_x^2+\sigma_y^2)  + \frac{1}{n+1}\sum_{t=0}^nM(t).
\end{align}

\begin{figure} 
\centerline{\includegraphics[scale=1]{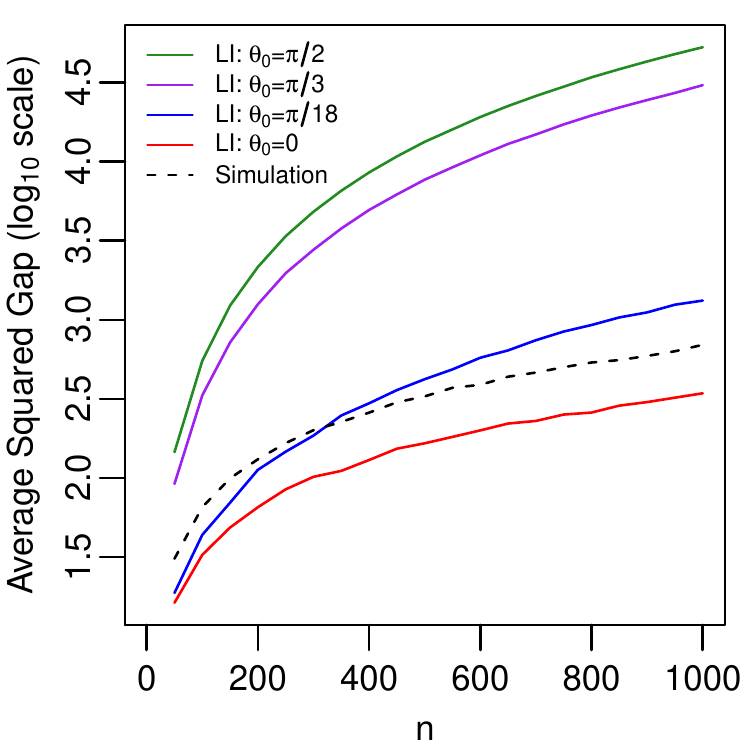}}
\caption{\textbf{Expected average gap between imputed trajectories and the true unobserved trajectory.} The length of the interval of missingness, $n$, ranges from $50$ to $1000$ by increments of $50$. For each value of $n$, the average over $1000$ simulations are used to estimate the average squared gap for the simulation approach. While for small $n$ and small $\theta_0$ linear interpolation can be a better approximation to the true trajectory,  asymptotically the simulation approach is better for any amount of curvature.} \label{SquaredGapFig}
\end{figure}

When comparing the expected gap between $\hat{{\b L}}(t)$ and ${\b L}(t)$ in Equation \eqref{EHt} and the expected gap between $\tilde{{\b L}}(t)$ and ${\b L}(t)$ in Equation \eqref{EFt}, only Equation \eqref{EFt} has both a $\sigma^2$ component as well as a component comprised of $\mu$. The $\sigma^2$ term in Equation \eqref{EFt} is exactly $1/2$ of that in Equation \eqref{EHt}, and while the second component involving $\mu$  disappears in the case where $t=0$, $t=n$, or when $\mu(i)$ is constant. In all other cases it can add considerably to the expected gap. Only when $\theta_0=0$ is $\mu(i)$ always constant. In this case we would expect the average squared distance between ${\b L}(t)$ and $\hat{{\b L}}(t)$ to be twice as large as the average squared distance between ${\b L}(t)$ and $\tilde{\bf L}(t)$. In other words, when the expected trajectory has no curvature (i.e., it is a straight line), then linear interpolation is the best approximation of the true trajectory. As the true trajectory gains curvature, $\hat{{\b L}}(t)$ becomes a closer approximation to the true trajectory and linear interpolation becomes increasingly inaccurate (see Figure \ref{SquaredGapFig}).

This result tells us that using simulated trajectories from the distribution of \emph{unobserved} flights leads to a better accuracy, on average, than using linear interpolation to fill in the missing data; the only exception to this is if the true unobserved trajectory happens to be a straight line. While this result is demonstrated on a model that assumes we are able to simulate from the distribution of unobserved flights, which is generally not possible since normally only the distribution of \emph{observed flights} is available, our goal is to come as close as possible to this scenario by borrowing information from the `closest' observed flights. As we elaborated above, `close' could mean temporally close (using TL kernel), spatially close (using GL kernel), or close in the sense of leveraging the periodicity of human behavior due to the circadian rhythm along with spatial closeness (using GLC kernel).

\subsection{Variability in the biases of mobility measure estimation}

Regardless of the approach used for imputing over the missing intervals in a person's mobility trace, there can still be substantial bias in the mobility estimates that are calculated from the imputed data. After all, each approach assumes a different model; linear interpolation assumes constant linear movement over a missing interval, TL assumes that flights and pauses that occur nearby in time come from the same distribution, GL assumes that flights and pauses that occur nearby in space come from the same distribution, and GLC assumes that flights and pauses that occur at the same time of day and at the same place come from the same distribution. These models each try to approximate the true nature of human mobility, but seldom will any of these models precisely hold true.

\begin{figure} 
\centerline{\includegraphics[scale=1]{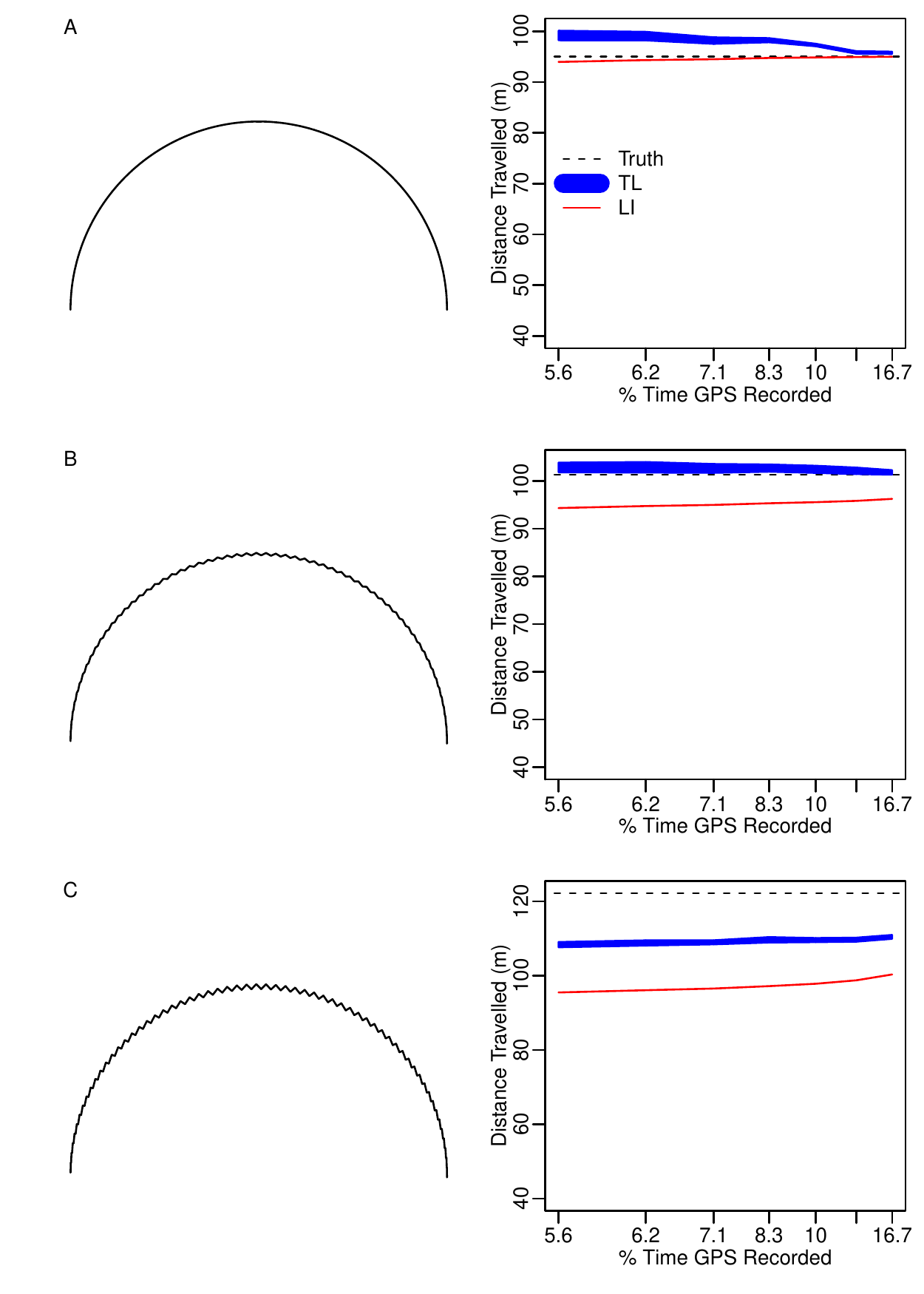}}
\caption{\textbf{Bias in the estimation of mobility metrics as the amount of missingness changes.} On the left, three different trajectories are displayed. The right of each panel shows the estimates of distance travelled for various levels of missingness in the trajectory on the left. Missingness is generated by taking evenly spaced intervals of different sizes (for different levels of missingness) out of the semicircular trajectories. For TL, each level of missingness is repeated for 100 simulated trajectories to obtain the $95\%$ confidence band. This is repeated for three different types of movement: a smooth trajectory (A), small jitters (B), and larger jitters (C). This demonstrate that both the direction and magnitude of a surrogate's bias in approximating the true trajectory can vary significantly depending on the true trajectory.} \label{6paneljitterFig}
\end{figure}

In addition, in most cases it is difficult to predict the direction of bias. We demonstrate this through an example by looking at an easily interpretable mobility measure: distance travelled. Consider a person who follows a semicircular trajectory with some added jitter to their movement (see Figure \ref{6paneljitterFig}). Evenly spaced intervals of different sizes are removed to show how the bias changes as the extent of missingness increases/decreases. As expected, in each case as missingness decreases, the bias in the estimates of distance travelled decreases. The bias is predictably negative for LI because LI takes the shortest possible path over missing intervals, so it attains the lower bound for the distance travelled metric. In contrast, the TL model is less predictable in the direction of its bias. For the smoother trajectories the TL approach overestimates distance travelled, but for a large enough jitter the bias switches direction and becomes negative. In this small toy example, GL will mirror TL in how the data is weighted, and there is no routine for GLC to take advantage of, so both GL and GLC are omitted here.

Overall, the $95\%$ confidence band of TL accurately reflects the amount of missingness in the data, visible as the narrowing of the confidence band as more and more data is observed, whereas the LI approach as a point estimator shows equally misplaced certainty regardless of the amount of missing data. The fact that the confidence bands do not in general attain the nominal $95\%$ coverage of the true distance travelled demonstrate the flaws, created by design in this example, of the TL assumption that nearby flights come from the same distribution. Unfortunately, when the majority of data is missing, it will be nearly impossible to avoid all bias when estimating various measures of mobility. Instead, one must choose a modeling assumption guided by domain specific knowledge that is as close to the truth as possible. To this end, in the next section, we compare various missing data imputation approaches across many measures of mobility in the context of empirical GPS data over a large sample of individuals.

\subsection{Mobility measure estimation on a week-long empirical mobility trace}

To generate a high-frequency GPS mobility trace, we had a test subject install an Android version of the Beiwe application on their phone for one week \citep{torous2016new}. The application was set to sample the smartphone GPS essentially in continuous time: the on-cycle was specified to be 119 minutes and the off-cycle just 1 minute. Ultimately, due to the occasional loss of power or GPS signal to their phone, an average of 92 minutes of GPS trajectories per day were missing as opposed to the expected 12 minutes per day, but the high quality of this data set allows us to establish a \textit{de facto} ground truth. The goal of this analysis is to take a subset of this data set and to simulate a higher rate of missingness, one that is likely observed in practice. We superimposed on top of the observed data a simulated 2-minute on-cycle and 10-minute off-cycle, and we calculated a variety of mobility measures on the data produced by multiple missing data imputation approaches (LI, TL, GL, and GLC). Here we report the error for each approach on the estimated mobility measures as compared to the ground truth.

\begin{figure} 
\centerline{\includegraphics[scale=.6]{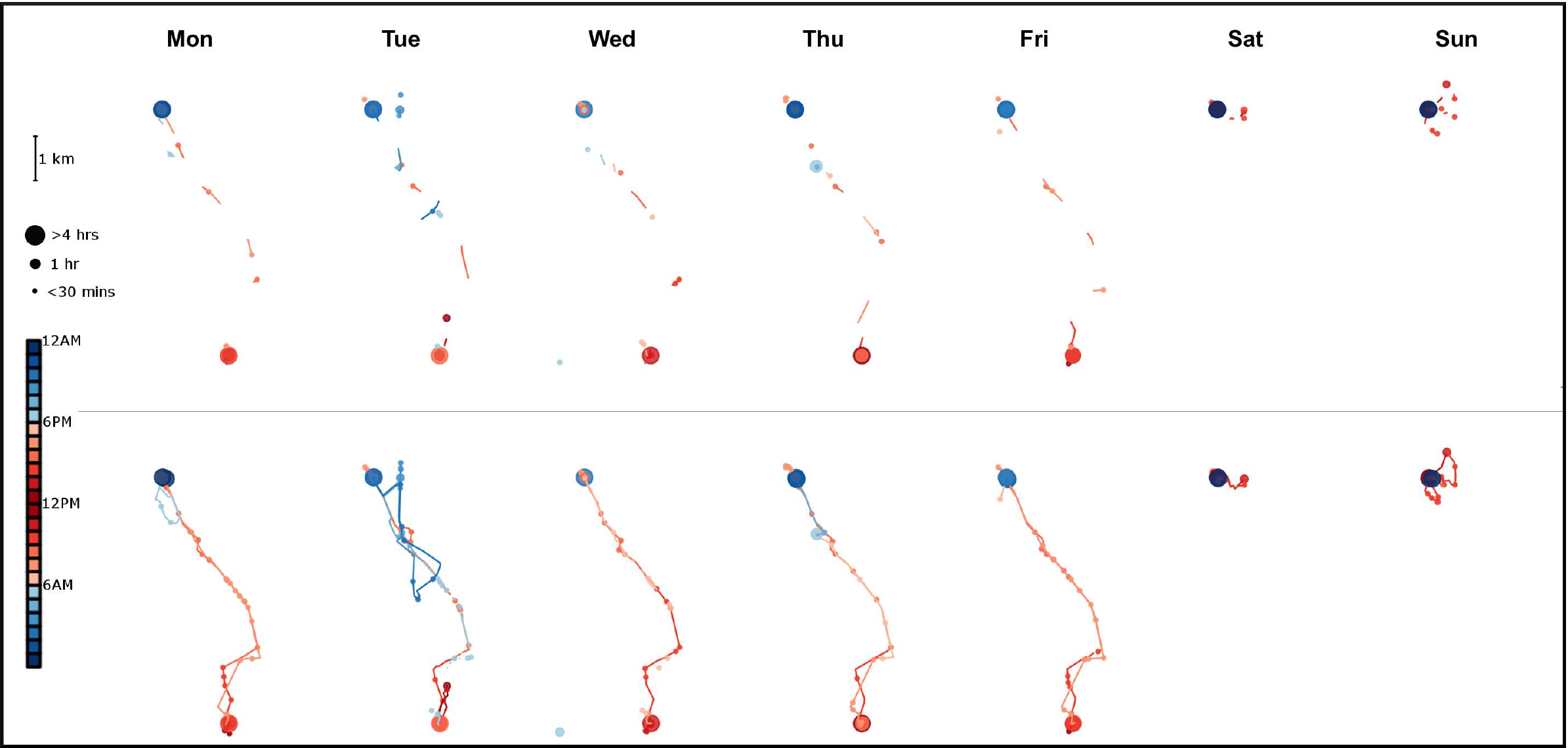}}
\caption{\textbf{A person's daily trajectories over the course of a week.} The bottom row represents a person's trajectory when GPS is captured continuously. The top row represents the identical trajectories to the bottom row with emulated/simulated missingness, such that the GPS is assumed to be recorded only for two-minute intervals with ten-minute gaps of missingness between recorded intervals. Lines represent flights, or movement. Points represent pauses, or periods where the person is stationary, with larger points indicating longer pauses.} \label{7panelFig}
\end{figure}

The person's daily mobility trace for the full week is displayed in Figure \ref{7panelFig}. The mobility trace based on the simulated 2-minute on-cycle and 10-minute off-cycle (top row) is shown alongside the mobility trace based on the complete data (bottom row). The general movements, locations, and daily routines are accurately captured by the subset with missingness, but some of the details are of course lost. For each day, $15$ different mobility measures were calculated (detailed in the Supplementary Materials), once for each missing data imputation approach and once for the ground truth. The estimates of the mobility measures for one example day are given in Table \ref{ExDailyFeaturesTbl}.

We also investigated sensitivity to changes in the cycle of data collection and planned missingness. Both 1-minute on-cycles and 2-minute on-cycles were paired with 10-minute off-cycles, 20-minute off-cycles, and 30-minute off-cycles. Absolute relative errors to the ground truth of our test subject for each mobility measure were calculated in of the six settings of planned data collection settings in Table \ref{RelAbsErrMisTbl}. While some measures, like radius of gyration and the probability of a pause, showed a clear decrease in accuracy as the missing interval was lengthened, some other measures, like the maximum distance from home and the number of significant locations visited, did not lose accuracy with increased missingness. In particular, the  1-minute on/10-minute off cycle showed a general improvement in accuracy relative to the 2-minute on/20-minute off cycle, despite the two different cycles having identical amounts of missingness.

\subsection{Analysis of the Geolife data set}

A larger sample of individuals is required for more generalized comparisons of the competing imputation methods, so we also used the complete data GPS trajectories from the Geolife data set for 855 outings across an additional 182 individuals  \citep{zheng2009mining,zheng2008understanding,zheng2010geolife}. Because only the mobility traces from user-specified outings are available, estimates of home and other significant locations are unreliable. As a result, we refrained from calculating those mobility measures that relate to significant locations for this data. Missingness was simulated according to the same 2-minute on-cycle and 10-minute off-cycle for each trajectory. The performance of each competing imputation approach was applied to the dataset with simulated missingness and evaluated against the ground truth. For each of the imputation approaches, TL, GL, and GLC, three different kernel parameter settings were considered. In each case $\nu=1$, while the scale parameter was varied (increased by a factor of $1$, $10$, and $20$). Increasing the scale parameter gave greater weight to nearby observations in resampling. The error was calculated by subtracting the estimated measure under each missing data imputation approach from that same measure calculated on the full data (with near-continuously gathered GPS). For the simulation-based imputation approaches, we used the mean value of the estimated measure from 100 simulated samples in the error calculations. 

A small error over most of the mobility measures would indicate that the resampling missing data imputation approaches (TL, GL, and GLC) do a good job of mimicking real human mobility patterns. To quantify this performance, the absolute value of the errors were averaged across all 855 outings for each mobility measure and for each imputation method (Table \ref{RelAbsErrTbl}). Based on this metric, the worst performing missing data imputation approach was linear interpolation (LI), with errors relative to the ground truth that were consistently larger than the resampling-based approaches for the majority of the mobility measures. The best performing missing data imputation approach for this data was TL with a scaling parameter of $20$, with nearly a 10-fold improvement in accuracy over LI.

\section{Discussion}
\label{sec4}
Past studies with small subject pools have not adequately accounted for missingness, likely because missing data is less of a problem for studies that provide their subjects with dedicated instrumentation capable of recording continuous or near-continuous GPS trajectories. However, instrumenting each subject with dedicated GPS devices is expensive and therefore scaling up to larger sample sizes or longer follow-up times becomes infeasible. In the near future, studies will likely increasingly leverage the high ownership rates of smartphones so that subjects need only to download an app onto their personal devices. For example, the smartphone research platform Beiwe is currently used to study  patient-centered outcomes across different disorders, from depression to surgical recovery, by collecting sensor data, survey data, and phone usage patterns from diverse patient cohorts. In these studies, which generally have long follow-up times, battery life is preserved by recording GPS less frequently, often leading to more than $80\%$ missing data. This means that missingness can no longer be ignored and will need to be properly adjusted for. In this paper, we introduced a hot-deck imputation approach to address missingness, and we found that, even with large percentages of missingness, mobility measure estimation from the proposed data imputation approach is accurate compared the current standard of using linear interpolation.

Our approach is designed to account for planned periods of missingness where the missing intervals may be frequent, but are each individually not too long. This type of missingness is benign because it operates independent of a person's location, and so can be treated as MCAR where imputation approaches like the one we propose are viable. While this planned missingness will undoubtedly account for the largest percentage of missing data in a person's GPS trace, other sources of missingness are left unaccounted for by our approach. If a person's position is obstructed from a satellite's view, it is possible that either a connection to satellite is not possible or their true location can be distorted. This type of missingness is difficult to account for and cannot be ignored as its mechanism qualifies as Missing Not at Random (MNAR) due to the missingness being dependent on location. Similarly, MNAR gaps in a person's mobility trace can be created by the person intentionally turning off their phone or disabling GPS.  Individuals that frequently have this type of missingness could potentially lead to large biases in the estimation of mobility measures. If an individual has a large amount of GPS data MNAR, which can be estimated by the extent of missingness there is outside of the scheduled intervals of missingness, the proposed imputation approach may not be appropriate.

With the prospect of scalable studies on the horizon, additional statistical challenges will likely emerge in the analysis of mobility measures from patient cohorts. With mobility measures paired with daily smartphone surveys, the longitudinal nature of the data can be leveraged with generalized linear mixed models \citep{breslow1993approximate} (GLMM) or generalized estimating equations \citep{liang1986longitudinal} (GEE) to estimate the effects of mobility measures on various outcomes obtained through the surveys. Also, while here we considered only mobility measures extracted from GPS traces, these mixed model frameworks can be readily adapted to include information from other smartphone sensors by adding additional covariates, such as those obtained from the phone's built-in accelerometer, into the regression model.

Finally, the method introduced in this paper has been implemented as a package in the statistical computing software, R, and is freely available (see Supplementary Materials). To conduct digital phenotyping studies, the Beiwe research platform can be used through its open source software.

\section{Supplementary Material}
\label{sec6}

The reader is referred to online Supplementary Materials for technical appendices, detailed feature definitions, and R software.

\section*{Acknowledgments}

{\it Conflict of Interest}: None declared.

\bibliographystyle{biorefs}
\bibliography{SimTrajectory-ref}

\newpage

\begin{table}
\caption{\label{ExDailyFeaturesTbl} \textbf{Mobility measures compared across different missing data imputation approaches for the trajectories of Figure 5.} GPS was collected continuously to establish the ground truth. For the missing data imputations, a Cauchy kernel was used with scale factor denoted by the number following the period. Larger scale factors give increased weight on nearby observations during resampling. For the TL, GL, and GLC approaches, the margin of error represents the standard deviation over 100 repeated simulations.}
\centering
\begin{tabular}{m{2.2cm}m{.9cm}m{.9cm}m{.9cm}m{.9cm}m{.9cm}m{.9cm}m{.9cm}m{1cm}m{1cm}m{.85cm}m{.85cm}}
\hline
\textbf{Measures}       & \textbf{TL.1}       & \textbf{TL.10}      & \textbf{TL.20}       & \textbf{GL.1}        & \textbf{GL.10}       & \textbf{GL.20}       & \textbf{GLC.1}       & \textbf{GLC.10}      & \textbf{GLC.20}      & \textbf{LI}  & \textbf{Truth}         \\
\hline
Hometime       & 831.5 $\pm$2.3     & 832.3 $\pm$2.4     & 833.4 $\pm$2.2     & 830.3 $\pm$2.2     & 830.5 $\pm$2.8     & 829.8 $\pm$1.9     & 829.1 $\pm$2.1    & 832.1 $\pm$2.2    & 831.3 $\pm$2.5     & 826.7   & 882.8   \\
\hline
DistTravelled  & 22184 $\pm$969.7 & 22446 $\pm$843.5 & 22569 $\pm$811.6 & 18801 $\pm$466.3 & 18801 $\pm$337.5 & 18779 $\pm$369.4 & 21791 $\pm$969.9 & 22380 $\pm$712.1 & 22444 $\pm$645.6 & 17236 & 19344 \\
\hline
RoG            & 2787.3 $\pm$2.3    & 2791.3 $\pm$2.6     & 2791.2 $\pm$1.9    & 2783.0 $\pm$1.6    & 2783.0 $\pm$1.9       & 2783.3 $\pm$2.5    & 2785.6 $\pm$1.3   & 2787.0 $\pm$1.5   & 2787.5 $\pm$1.8    & 2779.4  & 2781.3  \\
\hline
MaxDiam        & 6717 $\pm$169  & 6745 $\pm$129  & 6727 $\pm$98   & 6494 $\pm$44    & 6483 $\pm$8    & 6496 $\pm$34    & 6516 $\pm$55   & 6517 $\pm$55  & 6562 $\pm$94    & 6479  & 6467  \\
\hline
MaxHomeDist    & 6372 $\pm$165   & 6410 $\pm$123  & 6379 $\pm$93    & 6160 $\pm$49   & 6147 $\pm$16    & 6153 $\pm$39   & 6144 $\pm$30  & 6152 $\pm$5   & 6163 $\pm$24   & 6149  & 6129  \\
\hline
SigLocsVisited & 2.96 $\pm$0.73       & 3.20 $\pm$0.58        & 3.20 $\pm$0.71        & 3.16 $\pm$0.69       & 3.00 $\pm$0.76          & 2.96 $\pm$0.79       & 3.28 $\pm$0.61      & 3.12 $\pm$0.60       & 3.20 $\pm$0.65        & 2        & 3        \\
\hline
AvgFlightLen   & 172.7 $\pm$10.7    & 160.2 $\pm$7.6     & 158.6 $\pm$7.4     & 200.2 $\pm$23.2     & 193.2 $\pm$19.2    & 191.7 $\pm$18.1    & 129.9 $\pm$13.6   & 122.8 $\pm$6.1    & 127.1 $\pm$7.6     & 478.8   & 251.2   \\
\hline
StdFlightLen   & 152.9 $\pm$30.8    & 125.8 $\pm$10.1    & 123.2 $\pm$5.5     & 213.4 $\pm$51.5    & 205.8 $\pm$36.3    & 202.7 $\pm$43.5    & 151.0 $\pm$30.0   & 134.2 $\pm$8.4    & 137.1 $\pm$9.0     & 639.6   & 223.3   \\
\hline
AvgFlightDur   & 79.0 $\pm$9.3      & 69.4 $\pm$5.8      & 68.8 $\pm$5.6      & 119.0 $\pm$17.9    & 115.2 $\pm$13.4     & 113.5 $\pm$13.7    & 65.4 $\pm$10.5    & 57.2 $\pm$4.1     & 60.0 $\pm$5.1      & 340.6   & 77.0    \\
\hline
StdFlightDur   & 131.7 $\pm$17.0    & 115.3 $\pm$9.0     & 113.5 $\pm$10.2    & 170.3 $\pm$22.0    & 168.7 $\pm$14.8    & 166.7 $\pm$14.4   & 103.7 $\pm$18.2   & 85.0 $\pm$10.9    & 91.7 $\pm$13.1     & 289.8    & 55.2     \\
\hline
FracPause      & 0.88 $\pm$0.01          & 0.89 $\pm$0.01          & 0.89 $\pm$0.01          & 0.87 $\pm$0.01          & 0.87 $\pm$0.01          & 0.87 $\pm$0.01          & 0.87 $\pm$0.01         & 0.88 $\pm$0.01         & 0.88 $\pm$0.01          & 0.86     & 0.93     \\
\hline
SigLocEntropy  & 0.63 $\pm$0.01       & 0.63 $\pm$0.01       & 0.63 $\pm$0.01       & 0.63 $\pm$0.01          & 0.63 $\pm$0.01       & 0.63 $\pm$0.01          & 0.63 $\pm$0.01         & 0.63 $\pm$0.01         & 0.63 $\pm$0.01          & 0.63     & 0.63     \\
\hline
MinsMissing    & 1243       & 1243       & 1243      & 1243      & 1243       & 1243      & 1243     & 1243    & 1243      & 1243  & 92    \\
\hline
CircdnRtn      & 0.64 $\pm$0.02       & 0.63 $\pm$0.01       & 0.63 $\pm$0.02       & 0.67 $\pm$0.01       & 0.67 $\pm$0.01       & 0.67 $\pm$0.01       & 0.65 $\pm$0.02      & 0.66 $\pm$0.01      & 0.66 $\pm$0.02       & 0.69     & 0.66     \\
\hline
WkEndDayRtn    & 0.76 $\pm$0.02       & 0.76 $\pm$0.01       & 0.76 $\pm$0.01       & 0.78 $\pm$0.01       & 0.77 $\pm$0.01       & 0.78 $\pm$0.01       & 0.76 $\pm$0.02      & 0.76 $\pm$0.01      & 0.77 $\pm$0.01       & 0.81     & 0.79    \\
\hline
\end{tabular}
\hspace*{-1pc}
\end{table}

\begin{table}
\centering
\caption{\label{RelAbsErrMisTbl} \textbf{Comparison of imputation on different levels of missingness to the ground truth.} For each measure, the absolute error relative to the ground truth measure is scaled relative to the ground truth value as represented as a percent. For example, the value in the first cell contains $2.75$, indicating that using a cycle of 1 minute on and 10 minutes off for GPS missingness led to a $2.75\%$ error relative in the estimation of time spent at home to the ground truth. Over the one week period, absolute relative errors are calculated for each day using GLC with scaling paramter $10$ as the missing data imputation approach, and the median absolute relative error over the full week is shown. This is repeated for each mobility measure and each level of missingness. The Xon/Yoff indicates alternating regular intervals of X minutes of data collection as the on-cycle and Y minutes of scheduled missingness as the off-cycle.}

\begin{tabular}{m{2.2cm}m{1.5cm}m{1.5cm}m{1.5cm}m{1.5cm}m{1.5cm}m{1.5cm}}
\hline
\textbf{Measures}   & \textbf{1on/10off} & \textbf{1on/20off} & \textbf{1on/30off} & \textbf{2on/10off} & \textbf{2on/20off} & \textbf{2on/30off} \\
\hline
Hometime       & 2.75  & 6.29 & 7.53 & 2.18 & 5.58 & 5.83 \\
\hline
DistTravelled  & 15.02  & 16.06 & 24.59 & 8.59 & 22.44 & 21.75 \\
\hline
RoG            & 0.30  & 0.85 & 1.79 & 0.29 & 0.90 & 1.53 \\
\hline
MaxDiam        & 0.41  & 7.90  & 7.70 & 6.03 & 1.24 & 3.64 \\
\hline
MaxHomeDist    & 3.12  & 3.17 & 1.43 & 4.14 & 0.47 & 0.85 \\
\hline
SigLocsVisited & 32.14  & 11.11 & 21.43 & 23.21 & 23.21 & 0.00        \\
\hline
AvgFlightLen   & 63.92  & 63.54 & 56.28 & 30.86 & 51.96 & 60.51  \\
\hline
StdFlightLen   & 61.41  & 78.42 & 65.73 & 25.17 & 50.40 & 64.48 \\
\hline
AvgFlightDur   & 37.74  & 53.97 & 50.63 & 15.57 & 32.34 & 36.29 \\
\hline
StdFlightDur   & 55.69   & 228.24 & 562.80 & 112.36 & 40.70 & 530.51 \\
\hline
ProbPause      & 6.00  & 9.58 & 12.72 & 4.06 & 9.68 & 13.13 \\
\hline
SigLocEntropy  & 1.73   & 7.71 & 4.38  & 3.06 & 2.94 & 5.50 \\
\hline
CircdnRtn      & 1.27  & 6.86 & 5.41 & 0.97  & 6.85 & 6.91 \\
\hline
WkEndDayRtn    & 1.32  & 6.52 & 3.51 & 1.72 & 6.38 & 6.41 \\
\hline
\end{tabular}
\end{table}

\begin{table}[ht]
\centering
\caption{\label{RelAbsErrTbl}  \textbf{Comparison of different missing data imputation approaches to the ground truth.} $855$ outings/trajectories across 182 individuals from the Geolife data set were used to compare imputation approaches. $83.3\%$ missingness was imposed on the dataset in equal intervals. For each outing, all measures not requiring a home location or routine were calculated. For each measure, the error relative to the ground truth is stated as a percent. For the stochastic approaches, TL, GL, and GLC, the mean daily measures from 100 simulations are used in relative error calculations. The final row is the average absolute error across all measures. The best performing missing data approach, TL with a Cauchy kernel and scaling parameter of 20, is highlighted in blue. Highlighted in red, linear interpolation had the largest average error.}
\begin{tabular}{rrrrrrrrrrr}
  \hline
  & \textbf{LI} & \textbf{TL.1} & \textbf{TL.10} & \textbf{TL.20} & \textbf{GL.1} & \textbf{GL.10} & \textbf{GL.20} & \textbf{GLC.1} & \textbf{GLC.10} & \textbf{GLC.20} \\ 
  \hline
DistTravelled  & -1.44 & -0.15 & -0.26 & -0.58 & 1.70 & 0.20 & 0.08 & -0.67 & -0.83 & -0.71 \\ 
  \hline
 RoG  & -0.51 & 0.94 & 1.45 & 0.13 & 1.23 & 0.46 & 0.20 & 0.03 & -0.12 & 0.01 \\ 
  \hline
 MaxDiam  & -0.41 & 0.25 & 0.22 & -0.19 & 1.16 & 0.36 & 0.27 & -0.11 & -0.28 & -0.09 \\ 
  \hline
 AvgFlightLen  & 11.72 & -0.09 & -0.14 & -0.35 & 0.11 & 0.25 & 0.33 & 0.56 & 0.73 & 0.83 \\ 
  \hline
 StdFlightLen  & 10.62 & -0.11 & -0.03 & -0.65 & 0.56 & 0.65 & 0.85 & 1.07 & 1.03 & 1.29 \\ 
  \hline
 AvgFlightDur  & 22.55 & 0.55 & 0.40 & 0.47 & 0.12 & 0.62 & 0.69 & 1.51 & 1.64 & 1.73 \\ 
  \hline
 StdFlightDur  & 29.56 & 2.72 & 2.10 & 2.29 & 1.50 & 2.33 & 2.25 & 3.57 & 3.18 & 3.58 \\ 
  \hline
 ProbPause  & -10.01 & 5.22 & 5.36 & 3.80 & 10.26 & 7.88 & 7.05 & 5.35 & 4.66 & 4.38 \\ 
  \hline
 \textbf{Avg. Error}  & \cellcolor{red!20}10.85 & 1.26 & 1.24 & \cellcolor{blue!20}1.06 & 2.08 & 1.60 & 1.46 & 1.61 & 1.56 & 1.58 \\ 
   \hline
\end{tabular}
\end{table}

\end{document}